\documentclass[letterpaper,journal]{IEEEtran}

\usepackage{amsmath,amssymb,amsthm,mathtools}
\usepackage{bm}
\usepackage{cite}
\usepackage{booktabs}
\usepackage{array}
\usepackage[caption=false,font=normalsize,labelfont=sf,textfont=sf]{subfig}
\usepackage{textcomp}
\usepackage{stfloats}
\usepackage{placeins}
\usepackage{url}
\usepackage{graphicx}
\usepackage[dvipsnames]{xcolor}
\usepackage{tikz}
\usetikzlibrary{positioning,arrows.meta,shapes.geometric,fit,backgrounds}

\newtheorem{theorem}{Theorem}[section]

\newtheorem{assumption}[theorem]{Assumption}

\begin{document}

\title{Day-to-Day Traffic Network Modeling under Route-Guidance Misinformation:\
Endogenous Trust and Resilience in CAV Environments}

\author{Eunhan~Ka,~\IEEEmembership{Member,~IEEE},~and~Satish~V.~Ukkusuri,~\IEEEmembership{Senior~Member,~IEEE}%
\thanks{E.~Ka (ORCID: 0000-0003-0954-8075) and S.~V.~Ukkusuri
(ORCID: 0000-0001-8754-9925) are with the Lyles School of Civil and
Construction Engineering, Purdue University, West Lafayette, IN 47907 USA
(e-mail: kae@purdue.edu; sukkusur@purdue.edu). Corresponding author:
S.~V.~Ukkusuri.}%
\thanks{Manuscript received April 2026.}}

\markboth{Preprint - Under Review at IEEE TRANSACTIONS ON INTELLIGENT TRANSPORTATION SYSTEMS}%
{Ka and Ukkusuri: DTD Network Modeling under Route-Guidance Misinformation}

\maketitle

\begin{abstract}
Connected and autonomous vehicles and smart mobility services increasingly use digital route guidance as an operational input to traffic network management. When this information becomes unreliable or adversarial, day-to-day traffic models must represent not only flow adaptation but also the evolution of user trust in the information source. This paper develops a coupled day-to-day traffic assignment and trust-evolution framework for route-guidance misinformation. Within-day congestion is represented by Lighthill-Whitham-Richards network loading, while day-to-day route choice follows bounded-rationality logit learning with trust-dependent reliance on external guidance. Trust is modeled as an aggregate class-level behavioral reliance state encoded by a Beta evidence model and updated from repeated guidance errors. Theoretical analysis establishes stationary equilibria, a conservative stability guide, a weighted compliance index for population-level vulnerability, and an asymmetric recovery law that explains post-attack trust hysteresis. Numerical experiments on Sioux Falls, with an Anaheim robustness check, show that endogenous trust creates a threshold-based resilience mechanism. Below the trust-activation threshold, the attack remains behaviorally stealthy and dynamic trust provides almost no attenuation. Above the threshold, trust erosion reduces the impact of the fixed-trust attack by about 91 percent in Sioux Falls and 85 percent in Anaheim. The experiments also show that CAV penetration increases fixed-trust vulnerability while preserving dynamic attenuation, and that traffic performance can recover before trust, resulting in a 77-day hidden vulnerability window. The results provide a trust-aware modeling basis for resilience analysis in CAV-enabled traffic networks.
\end{abstract}


\section{Introduction}
\label{sec:introduction}

\IEEEPARstart{D}{igital} route guidance has become part of the operational layer of modern traffic networks. Navigation platforms influence daily route choices. Empirical and review evidence on advanced traveler information systems shows that real-time information changes route choice, learning, and compliance behavior~\cite{chorus2006atis,benelia2010route}. The role of information will grow as connected and autonomous vehicles (CAVs), vehicle-to-everything communication, and cloud-based routing become more common in smart mobility systems~\cite{bansal2017forecasting,talebpour2016cav}. This creates a modeling challenge for dynamic traffic assignment. The information channel that improves travel-time prediction and network management can also be manipulated, making the behavioral state of the traveler population part of the cyber-physical traffic system.

Empirical evidence shows that the threat is no longer hypothetical. Wang~et~al.~\cite{wang2018ghost} demonstrated that fabricated GPS traces can create phantom congestion in Waze. Eryonucu and Papadimitratos~\cite{eryonucu2022sybil} showed related Sybil-based manipulation on Google Maps, and S\"{o}derh\"{a}ll~et~al.~\cite{soderhall2025impact} reported that Sybil-based attacks on navigation mobile crowdsensing can increase average travel time by about 20 percent with Sybils representing 3 percent of the user population. GPS spoofing can also manipulate road navigation systems without user awareness~\cite{zeng2018gps}. Related CAV cybersecurity studies show that automated and connected vehicles introduce additional attack surfaces through sensing, communication, and cooperative-driving streams~\cite{petit2015cyberattacks,amoozadeh2015security}. These studies establish feasibility and immediate impact, but they do not explain how repeated misinformation changes route choice, trust, and network performance over multiple days.

Three streams of literature motivate the present framework. First, day-to-day (DTD) traffic assignment models describe how travelers learn from experience and how flows converge or oscillate over repeated route choices~\cite{cascetta1989dtd,smith1984stability,cantarella1995dtd}. Classical extensions considered idealized traveler information systems, stochastic stability, and link-based adjustment processes~\cite{friesz1994dtd,watling1999dtd,he2010dtd}. Recent doubly dynamic DTD studies integrate within-day network loading with day-to-day learning, imperfect information, bounded rationality, and information sharing~\cite{han2020dtd}. Advanced traveler information systems, stability, and mixed autonomy have also been examined in DTD settings~\cite{ye2021atis,iryo2020stabilisation,guo2022mixed,liang2023control}. These models usually treat information quality and information reliance as exogenous. CAV traffic studies further show that automation and connectivity can alter flow stability, capacity, and heterogeneous congestion patterns~\cite{talebpour2016cav,vandenberg2016bottleneck}, making trust-dependent reliance on information more important in mixed fleets. Second, route-guidance cybersecurity studies model attack mechanisms, static network effects, platform-side detection, or behavioral sensitivity to falsified travel-time information~\cite{lin2018dataintegrity,ka2025rga,yang2023anomaly,ukkusuri2025cyber}. They rarely represent multi-day behavioral adaptation with endogenous trust. Third, trust-in-automation research shows that trust is dynamic, learned from repeated outcomes, and asymmetric because failures reduce trust faster than successes rebuild it~\cite{lee2004trust,hoff2015trust,muir1996trust,dzindolet2003trust,kraus2020trust,guo2021trust}. Trust also plays a central role in the acceptance of automated vehicles~\cite{choi2015avtrust}. Beta reputation models provide a parsimonious way to encode binary success and failure evidence~\cite{josang2002beta,guo2021trust}. Driving simulator evidence further shows that multiple and potentially conflicting traffic information sources can affect route-choice behavior and source usage~\cite{imants2021conflicting}. These trust models, however, have not been embedded in dynamic traffic assignment.

This paper addresses the resulting gap: existing models do not jointly capture within-day congestion propagation, day-to-day flow adaptation, adversarial route guidance, and endogenous trust. The following research questions organize the study.

\begin{itemize}
\item \textbf{RQ1}: How does endogenous trust interact with day-to-day flow evolution under route-guidance misinformation?
\item \textbf{RQ2}: What determines the behavioral threshold at which trust-based attenuation activates, and what directional stability insights can be drawn from benchmark analysis?
\item \textbf{RQ3}: How does population composition, especially CAV penetration, govern network vulnerability?
\item \textbf{RQ4}: How do asymmetric trust erosion and recovery shape post-attack resilience?
\end{itemize}

Four contributions answer these questions. \textbf{C1} develops a coupled DTD and trust dynamical system and proves the existence of stationary regimes. \textbf{C2} derives a trust-activation guide and a conservative benchmark stability result that interprets the regimes of stealthy and detectable attacks. \textbf{C3} shows that a weighted compliance index, rather than population diversity alone, governs the leading vulnerability term and explains the dual effect of CAV penetration. \textbf{C4} establishes a post-attack trust recovery law and identifies a hidden vulnerability window in which traffic performance has recovered while trust remains depressed. These contributions are tested using LWR-based simulations on Sioux Falls, with Anaheim used as a larger-network robustness check. The rest of the paper is organized as follows. Section~\ref{sec:model} presents the model. Section~\ref{sec:theory} analyzes its theoretical properties. Section~\ref{sec:experiments} reports numerical experiments. Section~\ref{sec:conclusion} concludes.

\section{Model Formulation}
\label{sec:model}

\begin{figure}[!t]
\centering
\resizebox{\columnwidth}{!}{%
\begin{tikzpicture}[
  font=\footnotesize,
  node distance=5.2mm and 8.5mm,
  state/.style={draw, rounded corners=2pt, align=center, minimum width=26mm, minimum height=7mm, inner sep=2pt},
  process/.style={draw, align=center, minimum width=26mm, minimum height=7mm, inner sep=2pt},
  small/.style={draw, align=center, minimum width=25mm, minimum height=7mm, inner sep=2pt},
  arr/.style={-Latex, line width=0.45pt},
  darr/.style={-Latex, dashed, line width=0.45pt},
  lab/.style={font=\scriptsize, align=center}
]
\node[state] (flow) {Day $d$ flows\\$f^{(d)}$};
\node[process, below=of flow] (load) {Within-day\\LWR loading};
\node[state, below=of load] (cost) {Experienced costs\\$c(f^{(d)})$};
\node[process, below=9mm of cost] (update) {Perceived-cost\\update using $c,I,\lambda_k$};
\node[process, below=of update] (choice) {BR-MNL\\route choice};
\node[state, below=of choice] (next) {Day $d+1$ flows\\$f^{(d+1)}$};
\node[small, right=of cost] (info) {Guidance signal\\$I=c+a$};
\node[small, above=of info] (attack) {Attack injection\\$a_r(f)$};
\node[small, below=of info] (error) {Guidance error\\$e_k=|I-c|$};
\node[small, below=of error] (trust) {Beta trust update\\$(\alpha_k,\beta_k)\mapsto T_k$};
\node[small, below=of trust] (rely) {Guidance reliance\\$\lambda_k=\bar\lambda_kT_k$};
\draw[arr] (flow) -- (load);
\draw[arr] (load) -- (cost);
\draw[arr] (cost) -- (update);
\draw[arr] (update) -- (choice);
\draw[arr] (choice) -- (next);
\draw[arr] (next.west) -- ++(-4mm,0) |- node[lab, pos=0.83, left] {next-day\\feedback} (flow.west);
\draw[arr] (cost) -- (info);
\draw[darr] (attack) -- (info);
\draw[arr] (info) -- (error);
\draw[arr] (error) -- (trust);
\draw[arr] (trust) -- (rely);
\draw[arr] (rely.west) -- ++(-5mm,0) |- (update.east);
\begin{scope}[on background layer]
\node[draw, rounded corners=3pt, inner sep=4pt, fit=(flow)(load)(cost)(update)(choice)(next), label={[lab]above:Traffic state loop}] {};
\node[draw, rounded corners=3pt, inner sep=4pt, fit=(attack)(info)(error)(trust)(rely), label={[lab]above:Information and trust loop}] {};
\end{scope}
\end{tikzpicture}}
\caption{Coupled DTD and trust framework. The attack enters the information channel. Repeated guidance errors update trust, change reliance on guidance, and affect the next-day route-choice state through bounded-rationality multinomial logit (BR-MNL) route choice.}
\label{fig:framework}
\end{figure}
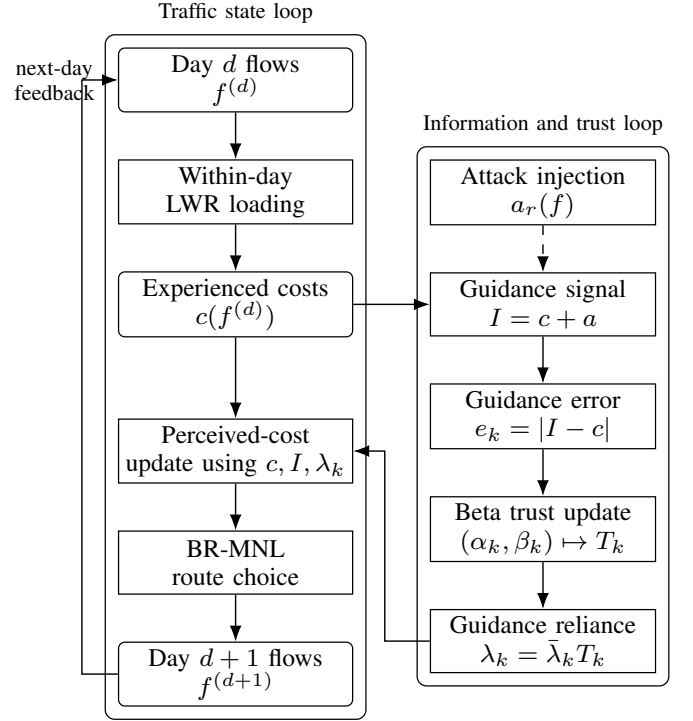

Fig.~\ref{fig:framework} summarizes the coupled DTD and trust framework. Day $d$ path flows generate experienced route costs through within-day loading. The information channel then reports a guidance signal, which may be manipulated during the attack window. The discrepancy between guidance and realized costs updates class-specific trust. Trust scales reliance on the information signal, and the resulting perceived costs determine the day $d+1$ route choice.

\subsection{Network, Loading, and Day-to-Day Learning}
\label{subsec:model-dtd}

Let $G=(\mathcal{N},\mathcal{A})$ denote a directed transportation network, $\mathcal{W}$ the set of origin-destination (OD) pairs, and $\mathcal{R}_w$ the path set for pair $w$. Demand $q_w$ is fixed over days. Driver classes are indexed by $k\in\mathcal{K}$ with population fractions $\pi_k$ and class-specific demand $q_w^k=\pi_k q_w$. The feasible path-flow set is
\begin{equation}
\mathcal{F}=\left\{f=\{f_r^k\}\ge 0:\sum_{r\in\mathcal{R}_w}f_r^k=q_w^k,\;\forall w,k\right\}.
\label{eq:feasible}
\end{equation}
For a day $d$ flow $f^{(d)}$, within-day dynamic network loading produces experienced path costs $c_r(f^{(d)})$. The numerical experiments use a route-only LWR kinematic-wave loading layer with Newell's simplified cumulative-curve propagation. The Sioux Falls and Anaheim network data and pre-enumerated path sets are based on the benchmark instances reported by Yu~et~al.~\cite{han2020dtd}. The theoretical analysis assumes only that the path-cost map is continuous on $\mathcal{F}$.

Drivers update perceived costs by blending personal experience and route guidance:
\begin{equation}
\begin{aligned}
\hat c_r^{k,(d+1)}={}&\lambda_m\hat c_r^{k,(d)}+(1-\lambda_m)\big[(1-\lambda_k^{(d)})c_r(f^{(d)})\\
&+\lambda_k^{(d)}I_r^{(d)}\big],
\end{aligned}
\label{eq:cost-update}
\end{equation}
where $\lambda_m\in(0,1)$ is the memory factor, $I_r^{(d)}$ is the guidance signal, and $\lambda_k^{(d)}$ is trust-dependent information reliance. Given perceived costs, class $k$ uses a bounded-rationality admissible set
\begin{equation}
\mathcal{B}_w^k(\hat c)=\left\{r\in\mathcal{R}_w:\hat c_r^k\le \min_{s\in\mathcal{R}_w}\hat c_s^k+\delta_k\right\},
\label{eq:br-set-v3}
\end{equation}
and a multinomial logit (MNL) response on that set:
\begin{equation}
P_r^k(\hat c)=
\frac{\exp(-\theta_k\hat c_r^k)\mathbf{1}\{r\in\mathcal{B}_w^k(\hat c)\}}
{\sum_{s\in\mathcal{B}_w^k(\hat c)}\exp(-\theta_k\hat c_s^k)}.
\label{eq:mnl-v3}
\end{equation}
The next-day class flow is $f_r^{k,(d+1)}=q_w^kP_r^k(\hat c^{k,(d)})$.

\subsection{Trust Dynamics and Information Reliance}
\label{subsec:model-trust-v3}

For each class $k$, trust is represented by a Beta evidence state $(\alpha_k^{(d)},\beta_k^{(d)})$ with expected trust
\begin{equation}
T_k^{(d)}=\frac{\alpha_k^{(d)}}{\alpha_k^{(d)}+\beta_k^{(d)}}.
\label{eq:trust}
\end{equation}
In this paper, $T_k^{(d)}$ is an aggregate class-level behavioral reliance proxy. It is not intended to represent a full cognitive model of individual trust formation.
The flow-weighted guidance error is
\begin{equation}
e_k^{(d)}=\frac{1}{Q_k}\sum_{w\in\mathcal{W}}\sum_{r\in\mathcal{R}_w}f_r^{k,(d)}\left|I_r^{(d)}-c_r(f^{(d)})\right|,
\label{eq:error}
\end{equation}
where $Q_k=\sum_w q_w^k$. The exact trust update used in all simulations classifies guidance as accurate if $e_k^{(d)}<\epsilon_k$ and inaccurate if $e_k^{(d)}>\epsilon_k$:
\begin{equation}
\Xi_k(e)=
\begin{cases}
\{1\}, & e<\epsilon_k,\\
[0,1], & e=\epsilon_k,\\
\{0\}, & e>\epsilon_k.
\end{cases}
\label{eq:xi}
\end{equation}
Given $\xi_k^{(d)}\in\Xi_k(e_k^{(d)})$, evidence evolves as
\begin{equation}
\begin{aligned}
\alpha_k^{(d+1)}&=\lambda_f\alpha_k^{(d)}+w_{s,k}\xi_k^{(d)},\\
\beta_k^{(d+1)}&=\lambda_f\beta_k^{(d)}+w_{f,k}(1-\xi_k^{(d)}).
\end{aligned}
\label{eq:trust-update}
\end{equation}
where $\lambda_f\in(0,1)$ is the forgetting factor and $w_{f,k}>w_{s,k}$ encodes faster trust erosion after failures than recovery after successes. The reliance coefficient in\eqref{eq:cost-update} is
\begin{equation}
\lambda_k^{(d)}=\bar\lambda_k T_k^{(d)}.
\label{eq:lambda}
\end{equation}
Thus, complete distrust implies exclusive reliance on personal experience.

Three classes are used in the experiments: a CAV-connected class with high information reliance and no bounded-rationality band, an app-reliant class, and an experience-based class. The default class shares are 10 percent, 60 percent, and 30 percent, respectively. These shares and trust parameters in Table~\ref{tab:class-params} are illustrative rather than field-calibrated. They encode distinct information-reliance profiles and vary in composition and sensitivity across experiments. Table~\ref{tab:class-params} reports the default behavioral parameters. For the CAV class, $T_k$ is interpreted as a composite reliance state that aggregates passenger acceptance of the routing system and perceived reliability of the vehicle-to-cloud information pipeline. The indifference-band magnitudes follow the scale of prior bounded-rationality DTD experiments~\cite{han2020dtd}, while the reliance and evidence weights encode high-, medium-, and low-reliance traveler profiles.

\begin{table}[!t]
\centering
\caption{Default class parameters used in the numerical experiments.}
\label{tab:class-params}
\footnotesize
\begin{tabular}{lccccc}
\toprule
Class & Share & $\delta_k$ & $w_{f,k}$ & $w_{s,k}$ & $\bar\lambda_k$ \\
\midrule
CAV-connected & 0.10 & 0 s & 0.8 & 0.16 & 0.9 \\
App-reliant & 0.60 & 200 s & 0.5 & 0.10 & 0.7 \\
Experience-based & 0.30 & 400 s & 0.3 & 0.06 & 0.3 \\
\bottomrule
\end{tabular}
\end{table}

\subsection{Route-Guidance Misinformation}
\label{subsec:model-threat-v3}

The paper focuses on a recommendation-layer misinformation attack, which perturbs displayed route costs or estimated times of arrival (ETAs). The manipulated signal is
\begin{equation}
I_r^{(d)}=c_r(f^{(d)})+a_r(f^{(d)}),\qquad
a_r(f)=-\gamma c_r(f)\omega_r,
\label{eq:attack}
\end{equation}
where $\gamma\in[0,1]$ is attack intensity and $\omega_r=1$ if route $r$ traverses at least one targeted link. The target set $\mathcal{A}_{\mathrm{att}}$ contains the top $N_{\mathrm{att}}$ links by topological betweenness centrality computed from public network topology. This attack is a stylized upper-bound stressor for trust-aware network modeling. It does not model exploit chains, stealth engineering, platform-side defenses, or attacks on the displayed candidate-route set.

\section{Theoretical Analysis}
\label{sec:theory}

This section gives the analytical backbone for the numerical results. The equilibrium and recovery results apply to general continuous path-cost maps. The stability and composition results are derived from two-link benchmarks and are used as directional guides for the full-network LWR simulations.

\begin{assumption}
\label{assumption-main-v3}
The path-cost map is continuous on $\mathcal{F}$; the attack perturbation map is continuous and bounded; the route-choice map is either continuous or upper hemicontinuous under graph completion; and the guidance-error map is continuous for each class.
\end{assumption}

\subsection{C1: Stationary Regimes}
\label{subsec:theory-c1-v3}

At stationarity, the perceived route cost for class $k$ is
\begin{equation}
z_r^k(f,T)=c_r(f)+\bar\lambda_k T_k a_r(f).
\label{eq:stationary-cost}
\end{equation}
For the exact threshold model, define the stationary trust selector
\begin{equation}
\Theta_k(f)=\left\{\frac{w_{s,k}\xi}{w_{s,k}\xi+w_{f,k}(1-\xi)}:\xi\in\Xi_k(e_k(f))\right\}.
\label{eq:theta-selector}
\end{equation}
Let $Q(f,T)$ denote the next-day flow correspondence induced by route choice under $z^k(f,T)$, and let $\Theta(f)=\prod_k\Theta_k(f)$.

\begin{theorem}[Existence of stationary equilibria]
\label{thm:c1-v3}
Under Assumption~\ref{assumption-main-v3}, the exact coupled DTD and trust system admits at least one stationary equilibrium $(f^*,T^*)\in\mathcal{F}\times[0,1]^K$ satisfying
\begin{equation}
(f^*,T^*)\in Q(f^*,T^*)\times\Theta(f^*).
\end{equation}
For a smooth trust regularization, existence follows from Brouwer's fixed-point theorem, and uniqueness follows whenever the smooth equilibrium map is contractive.
\end{theorem}

The proof uses graph completion for both the trust threshold and the bounded-rationality boundary, then applies Kakutani's fixed-point theorem. The result explains why the simulations settle into well-defined stealthy, degraded-trust, or post-attack recovery regimes.

\subsection{C2: Trust Activation and Stability Guide}
\label{subsec:theory-c2-v3}

The behavioral activation threshold follows directly from\eqref{eq:error} and\eqref{eq:attack}. At the pre-attack flow $f^{(0)}$, the first-day class-specific error under the primary recommendation-layer attack is
\begin{equation}
e_k^{(0)}=\gamma Q_k^{-1}\sum_w\sum_{r\in\mathcal{R}_w} f_r^{k,(0)}\omega_r c_r(f^{(0)}),
\end{equation}
which gives
\begin{equation}
\hat\gamma_k=\frac{\epsilon_k}{Q_k^{-1}\sum_w\sum_{r\in\mathcal{R}_w} f_r^{k,(0)}\omega_r c_r(f^{(0)})}.
\label{eq:gamma-hat}
\end{equation}
Thus, $\epsilon_k$ shifts the activation boundary. The accuracy-tolerance sensitivity in the Supplementary Material confirms that the first-day error scale is $D=536.05$~s on Sioux Falls and that changing $\epsilon$ shifts the error-based activation boundary while preserving the stealthy and detectable regimes under the tested tolerances.

For local stability, consider a smooth two-link benchmark with one OD pair, unit demand, $c_1(x)=c_0+bx$, $c_2(x)=c_0+b(1-x)$, and an attacked link with perturbation magnitude $\gamma$. Let $x_d=\Psi(h_d;\delta)$, where $h_d$ is the perceived cost difference and $L_\Psi(\delta)=\sup_h|\Psi_h(h;\delta)|$.

\begin{theorem}[Conservative sufficient stability bound]
\label{thm:c2-v3}
For the smooth two-link benchmark, suppose $2bL_\Psi(\delta)<1$. Define
\begin{align}
\gamma_1^*(\delta)&=\frac{1-2bL_\Psi(\delta)}{\bar\lambda\{bL_\Psi(\delta)+c_{\max}(1-\lambda_f)/w_s\}},\\
\gamma_2^*(\delta)&=\frac{1-\lambda_f}{w_fL_\rho bL_\Psi(\delta)},
\end{align}
and $\gamma^*(\delta)=\min\{\gamma_1^*(\delta),\gamma_2^*(\delta)\}$. Any stationary equilibrium is locally asymptotically stable whenever $0\le\gamma<\gamma^*(\delta)$. If $L_\Psi(\delta)$ is nonincreasing in $\delta$, then the guaranteed stability margin is nondecreasing in $\delta$.
\end{theorem}

Theorem~\ref{thm:c2-v3} is not a full-network stability theorem for the LWR simulations. It is a conservative benchmark result that explains why route-choice sensitivity can reduce stability margins and why bounded rationality can increase them.

\subsection{C3: Population Composition and Weighted Compliance}
\label{subsec:theory-c3-v3}

Let $u_k=\bar\lambda_kT_k^0$, $\Theta(\pi)=\sum_k\pi_k\theta_k$, and
\begin{equation}
\chi(\pi)=\sum_k\pi_k\theta_k u_k.
\label{eq:compliance}
\end{equation}
The scalar $\chi(\pi)$ is the weighted compliance index.

\begin{theorem}[Weighted compliance]
\label{thm:c3-v3}
For the attacked two-link logit benchmark, the equilibrium deviation from the symmetric split satisfies
\begin{equation}
x^*(\pi,\gamma)-\frac{1}{2}=-\frac{\gamma c^0\chi(\pi)}{4+2b\Theta(\pi)}+O(\gamma^2),
\end{equation}
where $c^0=c_0+b/2$. The first nontrivial relative total system travel time (TSTT) impact, or price of attack (PoAtt), is
\begin{equation}
\mathrm{PoAtt}(\pi,\gamma)=\frac{2bc^0}{(4+2b\Theta(\pi))^2}\chi(\pi)^2\gamma^2+O(\gamma^3).
\label{eq:poatt-leading}
\end{equation}
Thus, the leading vulnerability is governed by weighted compliance rather than by population diversity alone.
\end{theorem}

This result predicts that increasing the share of high-reliance CAV users increases fixed-trust vulnerability. Under dynamic trust, the same class also erodes trust the fastest, preserving attenuation once the trust threshold is crossed.

\subsection{C4: Recovery and Hidden Vulnerability}
\label{subsec:theory-c4-v3}

Suppose the attack ends at day $d_0$ and all guidance is accurate afterward. Let $n=d-d_0$, $A_k=w_{s,k}/(1-\lambda_f)$, and let $(\alpha_{k,0},\beta_{k,0})$ be the evidence state at attack termination.

\begin{theorem}[Post-attack trust recovery]
\label{thm:c4-recovery-v3}
Once the attack terminates,
\begin{equation}
\alpha_{k,n}=A_k+(\alpha_{k,0}-A_k)\lambda_f^n,\qquad
\beta_{k,n}=\beta_{k,0}\lambda_f^n,
\end{equation}
and
\begin{equation}
T_{k,n}=\frac{A_k+(\alpha_{k,0}-A_k)\lambda_f^n}{A_k+(\alpha_{k,0}+\beta_{k,0}-A_k)\lambda_f^n}.
\label{eq:recovery-law}
\end{equation}
If the attack phase is long and inaccurate, the trust recovery time grows as $\Theta(\log(w_{f,k}/w_{s,k}))$.
\end{theorem}

\begin{theorem}[Hidden vulnerability window]
\label{thm:c4-hidden-v3}
If the post-attack flow-perception subsystem is contractive and TSTT is Lipschitz in the flow vector, then TSTT recovers geometrically fast after attack cessation. If trust forgetting is slower than flow convergence, then there exists an interval in which TSTT has returned near baseline while trust remains below its pre-attack level.
\end{theorem}

The hidden window is a behavioral hysteresis effect. During this period, the network appears operationally recovered, but legitimate route guidance is less effective because users still discount the information source.

\subsection{Synthesis and Scope}
\label{subsec:theory-scope-v3}

The theory is intended to clarify mechanisms rather than to provide a closed-form solution for the multi-route LWR simulations. The equilibrium and recovery results apply to the general continuous-cost formulation. The stability and composition results are benchmarks that explain the directional effects of route-choice sensitivity, bounded rationality, and population compliance. The expected attenuation under dynamic trust can be summarized by an effective attack strength $S^{(d)}=\sum_k\pi_k\bar\lambda_kT_k^{(d)}\gamma$. If the excess PoAtt is locally quadratic in $S^{(d)}$, then
\begin{equation}
\begin{aligned}
\mathrm{PoAtt}_{\mathrm{dyn}}-1&\approx (\mathrm{PoAtt}_{\mathrm{fix}}-1)\eta(D),\\
\eta(D)&=\frac{1}{D}\sum_{d=0}^{D-1}\left(\frac{S^{(d)}}{S^{(0)}}\right)^2.
\end{aligned}
\label{eq:eta-v3}
\end{equation}
This relation is an interpretive approximation. In the simulations, the calibrated trust dynamics yield $\eta(50)\approx0.064$, which is consistent with the high trust-induced attenuation (TIA) observed after activation.

\section{Numerical Experiments}
\label{sec:experiments}

\subsection{Setup and Metrics}
\label{subsec:exp-setup-v3}

The primary network is Sioux Falls with 24 nodes, 76 links, 528 OD pairs, and 6,180 enumerated paths. Anaheim is used as a larger-network robustness check with 416 nodes, 914 links, 1,406 OD pairs, and 30,719 paths. Both networks use the benchmark network data and pre-enumerated path sets reported by previous studies~\cite{han2019computing,han2020dtd}. They are embedded here in the route-only LWR/Newell loading framework used for the misinformation experiments. The default parameters are $\lambda_m=0.7$, $\lambda_f=0.95$, $\epsilon=0.1$ h, $\theta=0.004$, $N_{\mathrm{att}}=10$, and an attack window from days 51 to 100 in a 200-day simulation. Daily TSTT is smoothed only for reporting by a six-day rolling average on Sioux Falls.

Attack impact is measured by the attack-window mean price of attack (PoAtt),
\begin{equation}
\mathrm{PoAtt}_{\mathrm{aw}}=\frac{\mathrm{mean}_{d\in\{51,\ldots,100\}}\mathrm{TSTT}(d)}{\mathrm{mean}_{d\in\{30,\ldots,50\}}\mathrm{TSTT}(d)}.
\label{eq:poatt-aw-v3}
\end{equation}
Trust-induced attenuation (TIA) is
\begin{equation}
\mathrm{TIA}=1-\frac{\mathrm{PoAtt}_{\mathrm{dyn}}-1}{\mathrm{PoAtt}_{\mathrm{fix}}-1},
\label{eq:tia-v3}
\end{equation}
reported only when the fixed-trust impact is above the numerical noise floor. Trust recovery time is the first post-attack day when mean trust exceeds 95 percent of its pre-attack value. The hidden vulnerability window is trust recovery time minus TSTT recovery time.

\begin{figure*}[t]
\centering
\IfFileExists{figures/exp_signature.png}{%
\includegraphics[width=.95\textwidth]{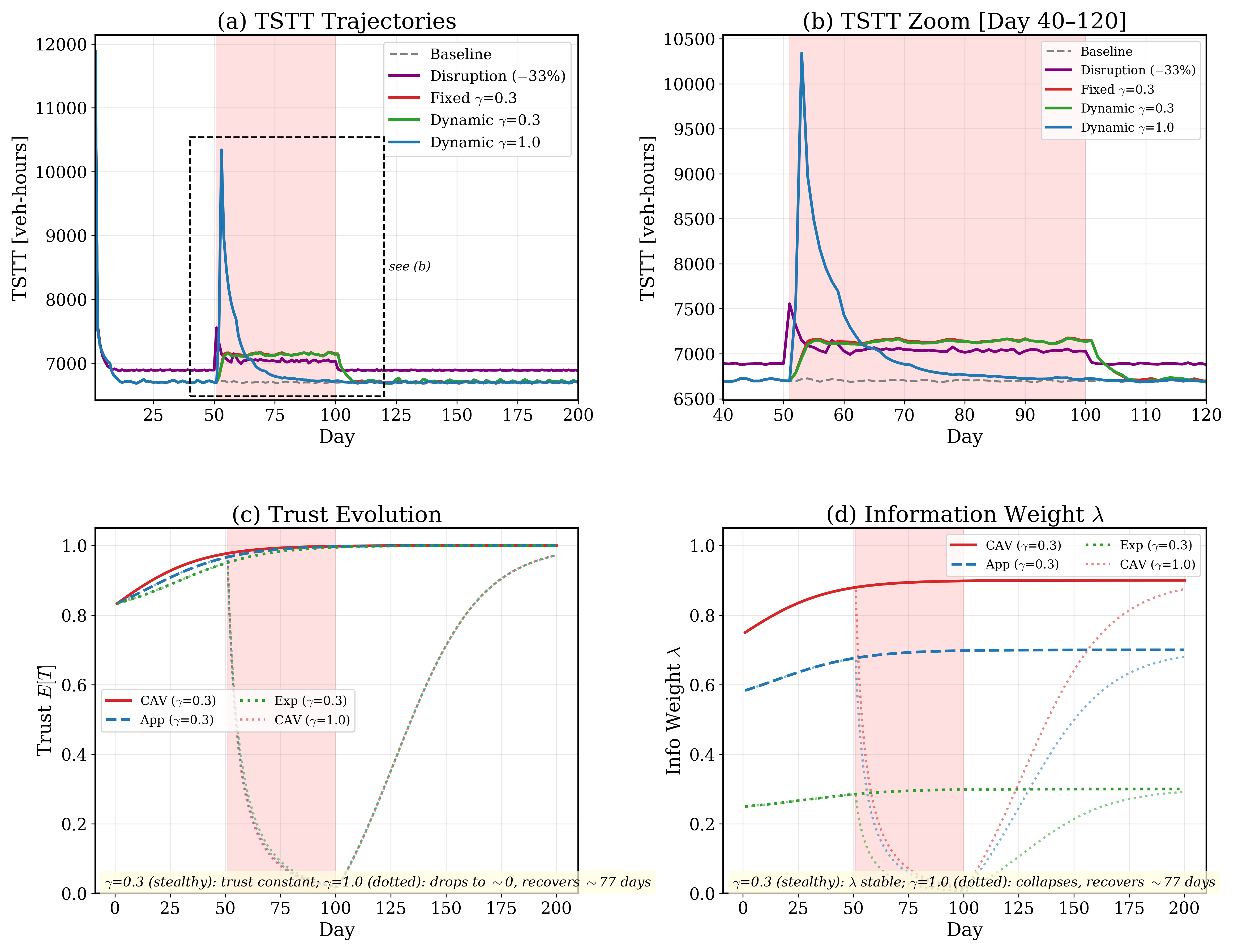}}{%
\fbox{\parbox{0.95\textwidth}{\centering Placeholder for \texttt{figures/exp\_signature.png}}}}
\caption{Signature scenarios on Sioux Falls under recommendation-layer route-guidance misinformation. Dynamic trust tracks fixed trust in the stealthy regime and attenuates the attack after the trust-activation threshold is crossed.}
\label{fig:signature}
\end{figure*}

\subsection{RQ1: Trust-Induced Resilience}
\label{subsec:exp-rq1-v3}

Fig.~\ref{fig:signature} shows the signature Sioux Falls scenarios. Under fixed trust with $\gamma=0.3$, recommendation-layer misinformation raises attack-window TSTT by 6.2 percent, with $\mathrm{PoAtt}_{\mathrm{aw}}=1.062$. Dynamic trust is nearly identical at this intensity, with $\mathrm{PoAtt}_{\mathrm{aw}}=1.060$ and TIA of about 2 percent. The reason is that guidance errors remain below the trust tolerance, so trust does not erode. This is the stealthy regime.

At higher intensity, the same mechanism changes qualitatively. For $\gamma=0.7$, fixed trust yields $\mathrm{PoAtt}_{\mathrm{aw}}=1.316$, while dynamic trust yields $1.028$, corresponding to about 91 percent attenuation. At $\gamma=1.0$, fixed trust reaches $1.653$, while dynamic trust remains near $1.056$. Trust erosion provides a behavioral resilience floor after the threshold is crossed, although the transient peak remains important. At $\gamma=1.0$, the peak PoAtt reaches about 1.541 before trust fully collapses.

The attack target comparison further shows that centrality-based targeting matters. At $\gamma=0.3$, topological betweenness centrality produces fixed-trust PoAtt 1.062, while a demand-aware path-betweenness benchmark produces 1.074. Random targeting yields $0.988\pm0.003$ over 10 seeds and is indistinguishable from the no-attack baseline. The two betweenness definitions overlap on only two of the ten target links, with a Pearson correlation $r=0.420$; however, both produce comparable disruption. This suggests that multiple backbone corridor sets can be disruptive once they interact with nonlinear congestion propagation.

\begin{figure*}[!t]
\centering
\IfFileExists{figures/exp_spatial_attack_sf.png}{%
\includegraphics[width=0.98\textwidth]{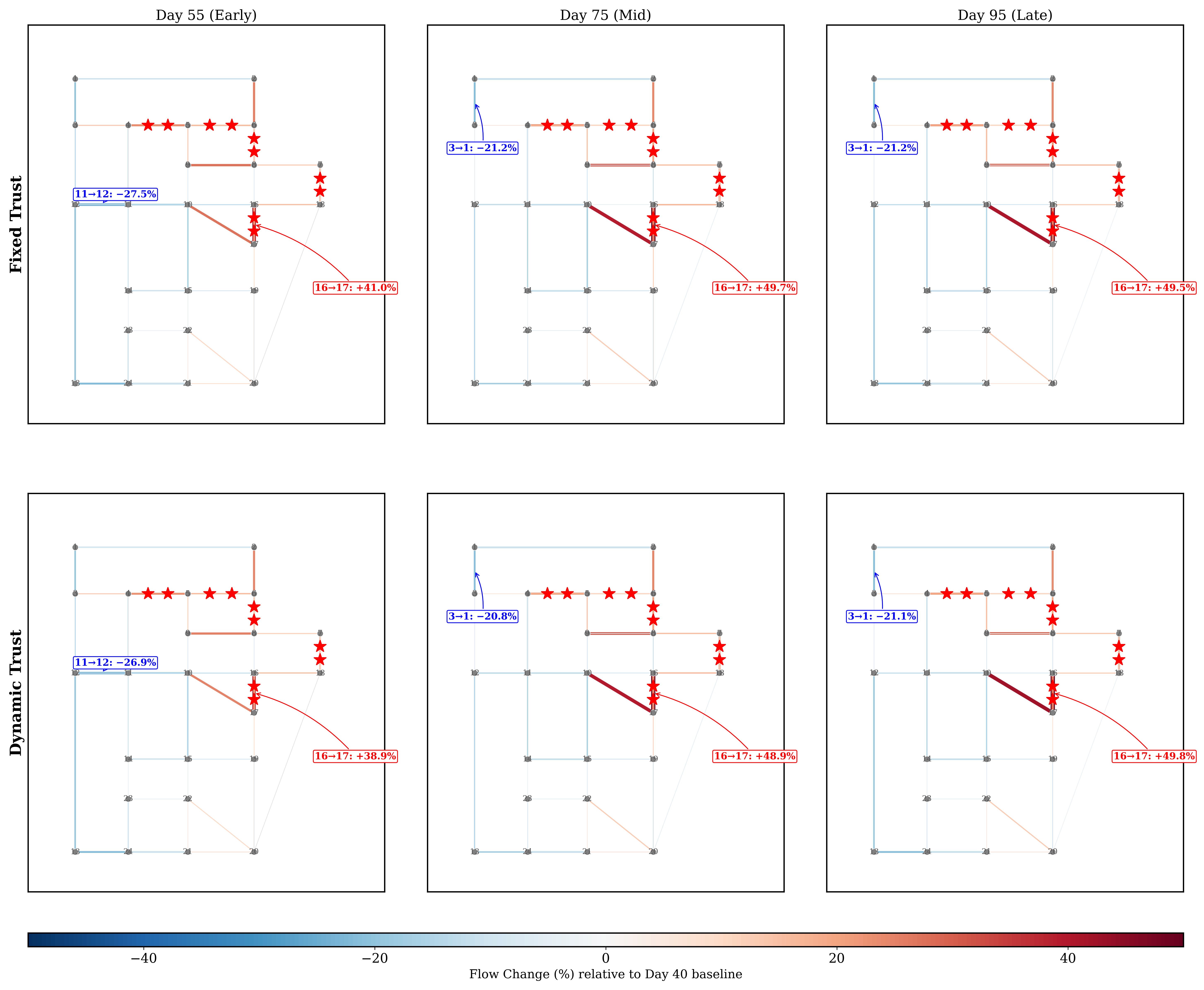}}{%
\fbox{\parbox{0.95\textwidth}{\centering Placeholder for \texttt{figures/exp\_spatial\_attack\_sf.png}}}}
\caption{Spatial flow change on Sioux Falls under recommendation-layer misinformation at $\gamma=0.3$. Link colors report the percentage change in flow relative to the Day~40 pre-attack baseline. Red star markers indicate the attacked links selected by topological betweenness centrality. Fixed and dynamic trust produce nearly identical patterns because trust does not erode in the stealthy regime.}
\label{fig:spatial-sf-main}
\end{figure*}

\begin{figure*}[!t]
\centering
\IfFileExists{figures/exp_spatial_attack_ana.png}{%
\includegraphics[width=\textwidth]{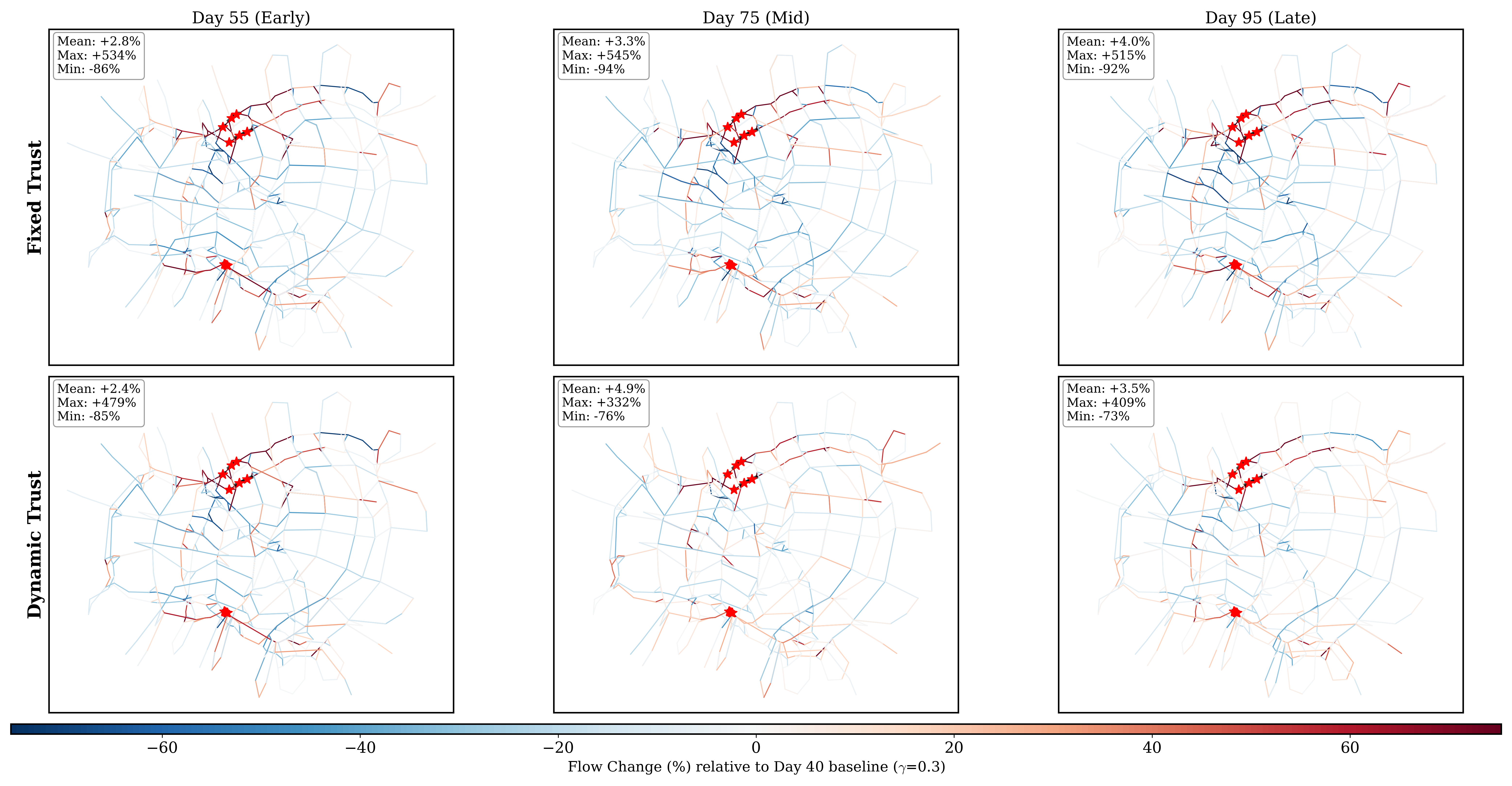}}{%
\fbox{\parbox{0.95\textwidth}{\centering Placeholder for \texttt{figures/exp\_spatial\_attack\_ana.png}}}}
\caption{Spatial flow change on Anaheim under recommendation-layer misinformation at $\gamma=0.3$. Link colors report the percentage change in flow relative to the Day~40 pre-attack baseline. Red star markers indicate the attacked links selected by topological betweenness centrality. The larger network shows more diffuse redistribution than Sioux Falls, consistent with lower target-path coverage and smaller fixed-trust impact.}
\label{fig:spatial-ana-main}
\end{figure*}

Figs.~\ref{fig:spatial-sf-main} and~\ref{fig:spatial-ana-main} provide spatial context for the targeted attack configuration. Red star markers indicate the attacked links, and link colors indicate the percentage change in flow relative to the Day~40 pre-attack baseline. In Sioux Falls, fixed and dynamic trust produce nearly identical redistribution patterns at $\gamma=0.3$, confirming that this intensity lies in the stealthy regime. In Anaheim, the same absolute attack budget affects a smaller fraction of the larger network, so the spatial response is more diffuse. This pattern is consistent with the lower fixed-trust impact and lower target-path coverage reported below.

Anaheim confirms the qualitative pattern on a larger network. At $\gamma=0.3$, fixed and dynamic PoAtt values are 1.029 and 1.016, giving TIA of about 46 percent. At $\gamma=0.7$, the corresponding values are 1.145 and 1.022, giving TIA of about 85 percent. Anaheim has lower target-path coverage than Sioux Falls, 22.6 percent compared with 62.0 percent, which explains its smaller fixed-trust impact. It also exhibits persistent DTD-LWR oscillations with a coefficient of variation near 1.2 percent, so it is used as a robustness check rather than as a precise calibration target. The cross-network result supports the two-regime mechanism without claiming exact quantitative transferability. The low-intensity Anaheim effect is modest relative to the endogenous DTD-LWR oscillation level, so the comparison is used as a qualitative robustness check rather than a precision calibration.

\begin{figure*}[!t]
\centering
\IfFileExists{figures/exp_anaheim.png}{%
\includegraphics[width=0.95\textwidth]{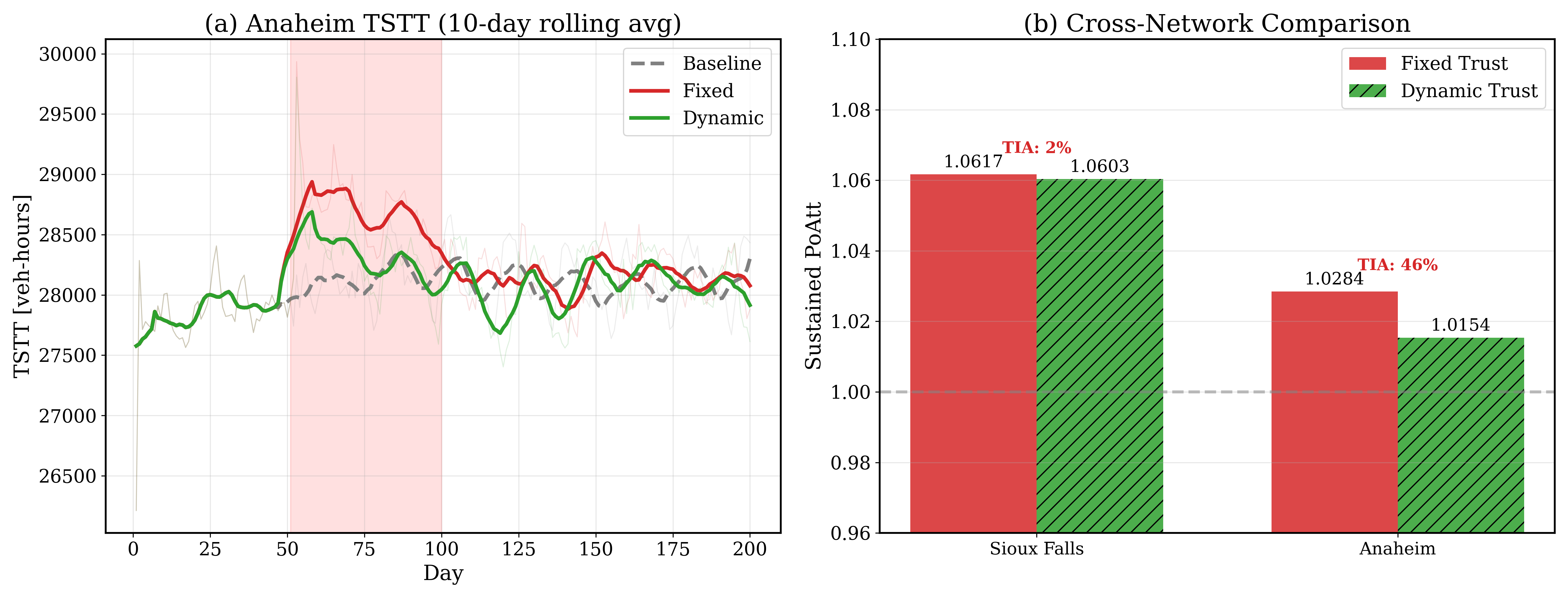}}{%
\fbox{\parbox{0.95\textwidth}{\centering Placeholder for \texttt{figures/exp\_anaheim.png}}}}
\caption{Anaheim robustness check. The larger network has lower target-path coverage than Sioux Falls and lower fixed-trust impact at $\gamma=0.3$, but dynamic trust still reduces the attack-window PoAtt.}
\label{fig:anaheim}
\end{figure*}

\subsection{RQ2: Trust Activation and Behavioral Parameters}
\label{subsec:exp-rq2-v3}

\begin{figure*}[!t]
\centering
\IfFileExists{figures/exp_gamma_sweep.png}{%
\includegraphics[width=.98\textwidth]{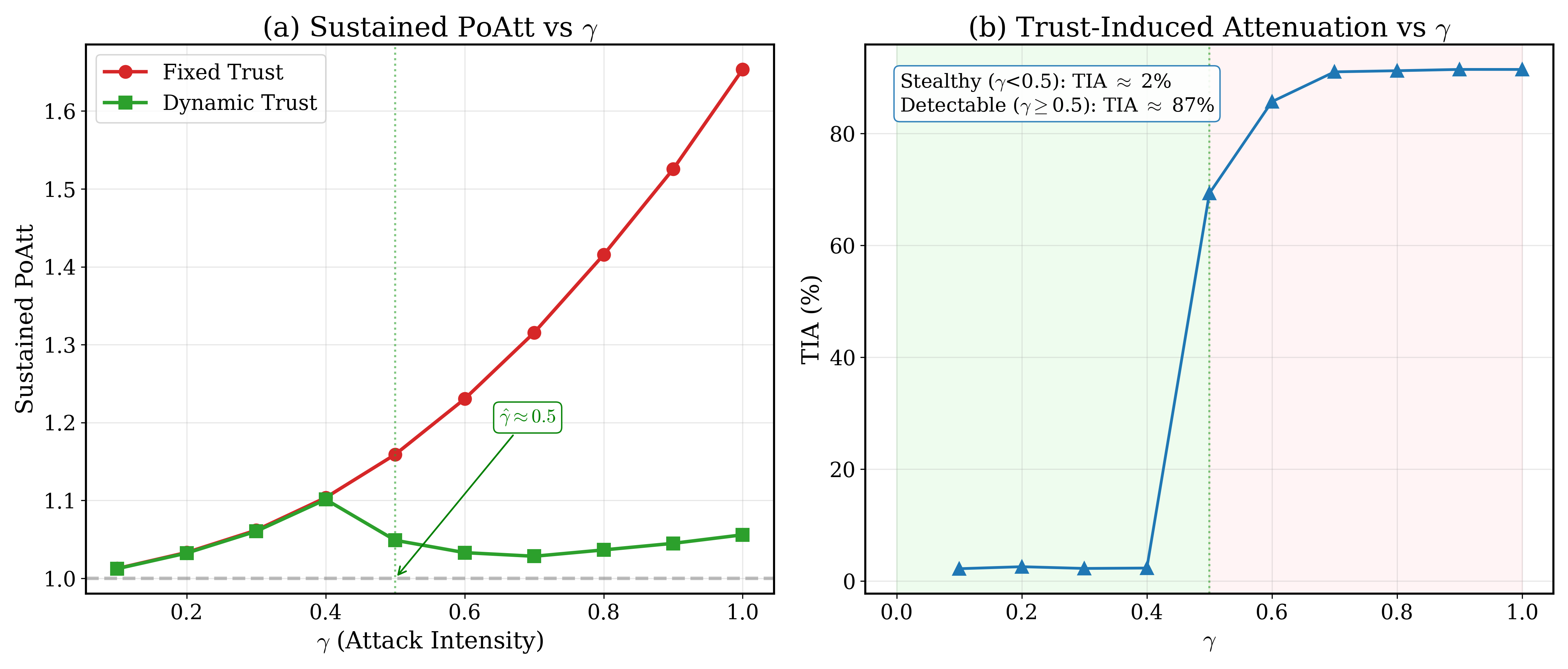}}{% 
\fbox{\parbox{0.9\textwidth}{\centering Placeholder for \texttt{figures/exp\_gamma\_sweep.png}}}}
\caption{Attack intensity sweep on Sioux Falls. The empirical trust-activation threshold is near $\hat\gamma=0.5$, separating stealthy and detectable regimes.}
\label{fig:gamma-sweep}
\end{figure*}

Fig.~\ref{fig:gamma-sweep} reports the intensity sweep. Under fixed trust, $\mathrm{PoAtt}_{\mathrm{aw}}$ increases monotonically from 1.013 at $\gamma=0.1$ to 1.653 at $\gamma=1.0$. Under dynamic trust, the curve follows the fixed-trust curve up to $\gamma=0.4$, then separates sharply at $\gamma\approx0.5$. At $\gamma=0.5$, dynamic trust reduces PoAtt from 1.159 to 1.049. For $\gamma\ge0.7$, dynamic PoAtt remains near 1.03 to 1.06 even as the fixed-trust impact continues to grow.

The threshold interpretation follows~\eqref{eq:gamma-hat}. A larger accuracy tolerance $\epsilon$ delays trust erosion, while a smaller tolerance activates it earlier. The accuracy-tolerance sensitivity in the Supplementary Material shows that the first-day error threshold scales with $\epsilon$ through $D=536.05$~s on Sioux Falls. The cumulative TIA-based regime boundary shifts consistently and preserves the stealthy and detectable regimes for $\epsilon\in\{0.05,0.10,0.15\}$~h. The empirical activation threshold $\hat\gamma\approx0.5$ should be interpreted alongside, rather than instead of, platform-side anomaly detection~\cite{yang2023anomaly}; behavioral resilience activates only after user-visible degradation becomes substantial.

Behavioral parameter sweeps support the conservative interpretation of Theorem~\ref{thm:c2-v3}. Bounded rationality changes the own-baseline fixed-trust PoAtt by less than 0.8 percentage points over $\delta\in[0,500]$ s at $\gamma=0.3$. At $\gamma=0.7$, TIA stays in the 89 percent to 92 percent range. A route-choice sensitivity sweep similarly shows that $\theta$ mainly shifts the pre-attack equilibrium TSTT rather than the relative attack impact under own-baseline normalization, so the benchmark analysis remains directionally consistent with the LWR simulations without claiming a full-network stability proof.

Two auxiliary sweeps are moved to the Supplementary Material because they support robustness rather than the main mechanism. Varying the information-sharing exponent changes fixed-trust PoAtt by less than 0.4 percentage points at $\gamma=0.3$ and by 2.7 percentage points at $\gamma=0.7$. The attack-budget sweep saturates near $N_{\mathrm{att}}=10$, indicating that the top-betweenness links already cover the main backbone corridors of Sioux Falls. These results indicate resilience redundancy: once trust dynamics dominate attenuation, memory-based information sharing has little additional effect.

\subsection{RQ3: Population Composition and CAV Penetration}
\label{subsec:exp-rq3-v3}

\begin{figure*}[!t]
\centering
\IfFileExists{figures/exp_composition_recovery_main.png}{%
\includegraphics[width=0.98\textwidth]{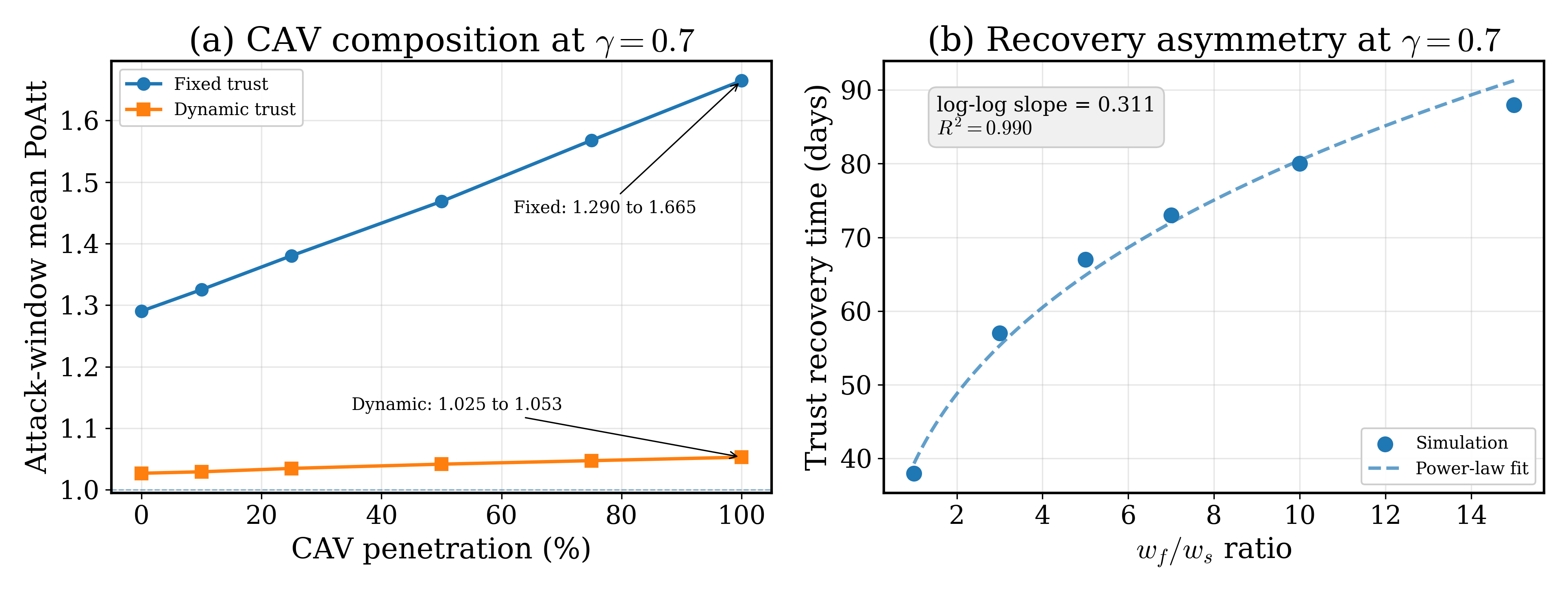}}{% 
\fbox{\parbox{0.9\textwidth}{\centering Placeholder for \texttt{figures/exp\_composition\_recovery\_main.png}}}}
\caption{Population composition and recovery asymmetry. (a) CAV penetration increases fixed-trust vulnerability but leaves dynamic-trust PoAtt near baseline after trust activation. (b) Trust recovery time increases sublinearly with the erosion-to-recovery ratio at $\gamma=0.7$, with log-log slope 0.311 and $R^2=0.990$.}
\label{fig:composition-recovery}
\end{figure*}

Fig.~\ref{fig:composition-recovery}(a) summarizes the composition sweep at $\gamma=0.7$ and confirms the weighted-compliance prediction. Under fixed trust, increasing the CAV share from 0 percent to 100 percent raises $\mathrm{PoAtt}_{\mathrm{aw}}$ from 1.290 to 1.665. This occurs because the CAV class has the highest information reliance and no bounded-rationality band. Under dynamic trust, the same sweep raises PoAtt only from 1.025 to 1.053, while TIA remains near 91 percent. CAV penetration has a dual effect. It increases vulnerability when users continue to trust compromised guidance, but it also concentrates trust erosion in the most information-reliant class after the detection threshold is crossed.

The benchmark scaling established in Theorem~\ref{thm:c3-v3} is further corroborated by the accompanying quantitative results. Using the composition sweep in Fig.~\ref{fig:composition-recovery}(a) and preserving the default 2:1 split between the app-reliant and experience-based classes within the non-CAV remainder, the normalized compliance term $\sum_k \pi_k \bar\lambda_k$ rises from 0.567 at 0 percent CAV to 0.900 at 100 percent CAV. Because $\chi(\pi)$ is proportional to this term under the common pre-attack trust and route-choice sensitivity used in the sweep, the leading-order excess-impact term in~\eqref{eq:poatt-leading} is expected to scale by about $(0.900/0.567)^2=2.52$. The observed excess fixed-trust PoAtt rises from 0.290 to 0.665 above baseline, a factor of 2.29, which is directionally consistent with the benchmark prediction.

This result should not be interpreted as a platform-side defense. It is a behavioral attenuation mechanism. In practice, this implies that CAV guidance systems require trust-aware fallback and anomaly-detection protocols to prevent high connectivity from becoming a persistent vulnerability.

\subsection{RQ4: Recovery and Hidden Vulnerability}
\label{subsec:exp-rq4-v3}

Fig.~\ref{fig:composition-recovery}(b) summarizes the post-attack recovery experiments. The results are consistent with the asymmetry predicted by Theorems~\ref{thm:c4-recovery-v3} and~\ref{thm:c4-hidden-v3}. At $\gamma=0.7$, increasing $w_f/w_s$ from 1 to 15 raises trust recovery time from 38 days to 88 days. A log-log fit gives a slope of 0.311 with $R^2=0.990$, consistent with sublinear recovery and the logarithmic bound in Theorem~\ref{thm:c4-recovery-v3}. Faster trust erosion also reduces dynamic-trust PoAtt during the attack window, from 1.078 at $w_f/w_s=1$ to 1.012 at $w_f/w_s=15$. At $\gamma=0.3$, the same asymmetry sweep has almost no effect because the attack remains below the trust-activation threshold and trust does not erode.

At $\gamma=1.0$, traffic performance returns to baseline immediately after the attack ends, while trust requires about 77 days to recover. The result is a hidden vulnerability window. During this interval, aggregate TSTT suggests that the network is healthy, but legitimate guidance has become less effective because users continue to discount the information source. This window is especially relevant for incident management because reliable post-event guidance may be needed precisely when trust has not yet recovered. The 77-day value refers to the all-class recovery metric under the $\gamma=1.0$ dynamic-trust scenario. The supplementary accuracy-tolerance analysis reports class-averaged recovery measures at $\gamma=0.7$, so those values are not numerically comparable to the main hidden-window metric.

\section{Conclusions}
\label{sec:conclusion}

This paper introduced a coupled DTD and trust-evolution framework for traffic network modeling under route-guidance misinformation in CAV and smart mobility environments. The central finding is that endogenous trust is a threshold-dependent mechanism of behavioral resilience. It does not protect the network below the trust-activation threshold, but it strongly attenuates the impact of attacks once repeated guidance errors become large enough to erode trust. The Supplementary Material provides compact traceability, accuracy-tolerance sensitivity, extended numerical checks, and proofs of the main theoretical results.

The practical implication is twofold. First, low-intensity misinformation can remain behaviorally stealthy, so platform-side anomaly detection should monitor guidance residuals before they exceed user tolerance. Second, aggregate traffic recovery is not enough to declare system recovery because trust may remain depressed for weeks. Compliance rates, route-following changes, user complaints, and ETA residuals should therefore be monitored alongside TSTT or delay.

Several limitations define the scope of the results. The recommendation-layer attack is a stylized upper-bound stressor and does not model exploit chains, platform detection, or a limited set of candidate route displays. The threshold trust model produces clear regimes but simplifies graded human responses. The Beta-based update is an aggregate behavioral proxy, not a full cognitive model of individual trust formation. The trust parameters are literature-informed and simulation-parameterized rather than estimated from field data. The stability and composition theorems provide benchmark-based analytical guidance, not full-network proofs. Finally, only Sioux Falls and Anaheim are tested. Future work should estimate trust dynamics from field or simulator data, model source-specific trust, add game-theoretic attacker and defender adaptation, and integrate trust-aware route guidance with platform-side anomaly detection. Endogenous trust should be viewed as a resilience floor, not as a substitute for active cybersecurity defenses.

\section*{Acknowledgment}
This work is based on research supported by the National Center for Transportation Cybersecurity and Resiliency (TraCR), a U.S. Department of Transportation National University Transportation Center headquartered at Clemson University, Clemson, SC, USA. The opinions, findings, and conclusions are those of the authors and do not necessarily reflect the views of TraCR or the U.S. Government.

\section*{Data and Code Availability}
The simulation code, network data, experiment configurations, and reproduction scripts are publicly available at \url{https://github.com/eunhanka/DTDRGA}.

\IEEEtriggeratref{24}
\bibliographystyle{IEEEtran}
\bibliography{reference_DTD}

\enlargethispage{2\baselineskip}
\begin{IEEEbiography}[{\includegraphics[width=1in,height=1.25in,clip,keepaspectratio]{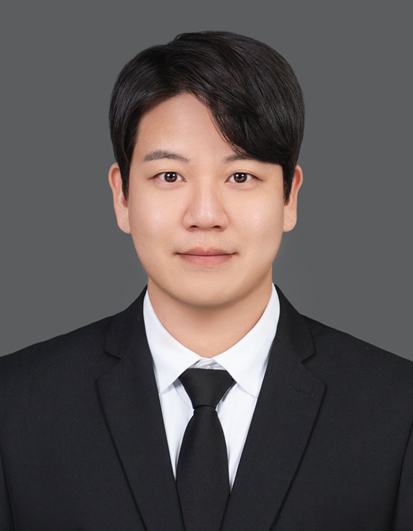}}]{Eunhan Ka}
is a Postdoctoral Researcher with the Lyles School of Civil and Construction Engineering, Purdue University. He received the B.S. and M.S. degrees in civil and environmental engineering from Seoul National University, Seoul, South Korea, in 2016 and 2018, respectively, and the Ph.D. degree in civil engineering from Purdue University, West Lafayette, IN, USA, in 2025. His research interests include transportation network modeling, traffic state estimation, transportation cybersecurity, network resilience, connected and autonomous mobility, and physics-informed and data-driven learning for transportation systems.
\end{IEEEbiography}

\begin{IEEEbiography}[{\includegraphics[width=1in,height=1.25in,clip,keepaspectratio]{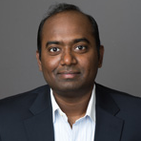}}]{Satish V. Ukkusuri}
is the Hubert and Audrey Kleasen Professor in Civil Engineering with the Lyles School of Civil and Construction Engineering and a Professor of Computer Science (courtesy) with Purdue University, West Lafayette, IN, USA. He received the B.Tech. degree in civil engineering from the Indian Institute of Technology Madras, Chennai, India, in 2001, the M.S. degree in civil engineering from the University of Illinois Urbana-Champaign, Urbana, IL, USA, in 2002, and the Ph.D. degree in civil engineering from The University of Texas at Austin, Austin, TX, USA, in 2005. His research interests include transportation network modeling, smart mobility, infrastructure resilience, connected and autonomous transportation systems, transportation cybersecurity, and data-driven urban analytics.
\end{IEEEbiography}

\end{document}